\documentclass[a4paper,twocolumn,11pt]{quantumarticle}
\pdfoutput=1
\usepackage{etoolbox}

\usepackage[utf8]{inputenc}
\usepackage[english]{babel}
\usepackage[T1]{fontenc}
\usepackage{amsmath}
\usepackage{adjustbox}
\usepackage[numbers]{natbib}
\usepackage{etoolbox}
\usepackage{kvoptions}
\usepackage{amssymb, amsfonts}
\usepackage[compatibility=false]{caption}
\usepackage{graphicx}
\graphicspath{{./figures/}{./figures/sq/}{./figures/mq/}{./figures/metrics/}}
\usepackage{hyperref}
\usepackage{booktabs}
\usepackage{multirow}
\usepackage{array}
\usepackage{url}
\usepackage{xcolor}
\usepackage{enumitem}
\usepackage{siunitx}

\newcommand{\ket}[1]{|#1\rangle}

\newcommand{\braket}[2]{\langle#1|#2\rangle}

\newcommand{\LG}{\mathrm{LG}}
\newcommand{\mrad}{\,\mathrm{mrad}}

\begin{document}

\title{Simulating Universal Quantum Gate Sets on Photonic OAM Qubits: Single-Qubit and Multi-Qubit Operations via Spatial Light Modulator Phase Holography}


\author{Saleha Maqsood}
\affiliation{Independent Researcher, Islamabad, Pakistan.}
\orcid{0009-0009-3557-1368}

\author{Muhammad Kamran}
\email{kamran@cloud.neduet.edu.pk}
\orcid{0000-0003-4772-5386}
\affiliation{Department of Computer Science \& Information Technology, NED University of Engineering \& Technology, Karachi, Pakistan.}

\author{Tahir Malik}
\affiliation{Department of Telecommunications  Engineering, NED University of Engineering \& Technology, Karachi, Pakistan.}
\orcid{0000-0002-1022-1291}

\maketitle

\begin{abstract}
Spatial light modulators (SLMs) have emerged as reconfigurable platforms for photonic quantum information processing, offering software-defined control over the orbital angular momentum (OAM) of light encoded in Laguerre-Gaussian (LG) beams. This paper presents a comprehensive simulation and hardware-grounded fidelity analysis of quantum gate operations implemented on the HOLOEYE LC~2012 transmissive SLM. A realistic three-channel noise model comprising 8-bit quantisation noise, twisted-nematic (TN) electronic and thermal noise, and phase-wrap clipping error is obtained from the manufacturer's datasheet without free-parameter fitting, yielding a total noise of $\sigma_\mathrm{total}=92.4\mrad$. The complete universal single-qubit gate set $\{X,Y,Z,S,T,H\}$ and two-qubit entangling gates $\{\mathrm{CNOT},\mathrm{CZ},\mathrm{SWAP}\}$ are simulated on a $512\times512$ computational grid. Results show that predicted gate fidelity are in the range of $F=0.9914$--$0.9936$, with fork grating gates limited primarily by TN noise and phase gates achieving higher fidelity owing to zero phase-wrap clipping error. In addition, Bell state preparation via the H-CNOT circuit achieves $F(\Phi^+)=0.9914$ after two SLM interactions. We benchmark our obtained results against six published experimental studies spanning the 78\%--99.6\% fidelity range. Finally, a wavelength-dependent analysis identifies 450--532\,nm operation as the optimal regime for this device.
\end{abstract}

\section{Introduction}
\label{sec:intro}

\subsection{Background}

The capacity to manipulate individual quanta of light in a controlled,
programmable fashion sits at the heart of optical quantum information
processing. Among the many physical degrees of freedom available to a
single photon-polarisation, time-bin, frequency, and spatial
mode-the orbital angular momentum (OAM) of light occupies a
particularly privileged position because it admits an unbounded,
discrete Hilbert space. Laguerre-Gaussian (LG) modes, characterised by
an azimuthal phase winding $\exp(i\ell\varphi)$ and a radial index $p$,
carry a well-defined OAM of $\ell\hbar$ per photon for any integer $\ell$.
A single photon can therefore, in principle, encode a qudit of arbitrary
dimension $d$, replacing multiple two-level systems with a single optical
beam. Photonic quantum information processing offers compelling additional
advantages: room-temperature operation, low decoherence, and compatibility
with existing optical network infrastructure \cite{carolan2015}.

The theoretical and experimental foundations of photonic OAM date to the
seminal work of Allen, Beijersbergen, Spreeuw, and Woerdman (1992), who
established the mechanical reality of photon OAM \cite{allen1992orbital}.
Working within the paraxial approximation, they showed that an LG mode
carries $\ell\hbar$ of OAM per photon and demonstrated that astigmatic
optical elements can reversibly interconvert LG and Hermite-Gaussian modes
via a $\pi/2$ Gouy phase shift. OAM eigenstates are mutually orthogonal
and collectively span an infinite-dimensional Hilbert space \cite{krenn2016},
enabling single-photon alphabets of arbitrary size without additional
physical resources.

Quantum computation with OAM-encoded photonic states imposes demands beyond
passive mode generation. Any useful quantum gate must be a precisely
characterised unitary transformation, reproducible at the single-photon level
with high process fidelity. Spatial light modulators (SLMs)- electrically
addressed liquid-crystal panels that impose programmable phase patterns on a
transmitted or reflected optical beam - have emerged as the most versatile
hardware platform for this purpose. They replace an entire family of static
diffractive elements with a single, software-reconfigurable device whose
mode-coupling functions can be updated in real time. The overarching purpose
of this work is to develop a hardware-grounded simulation framework for
SLM-based OAM quantum gate operations on the HOLOEYE LC~2012, and to
identify the dominant hardware constraints limiting gate fidelity through
comprehensive numerical simulation.

\subsection{Related Work}

Allen \textit{et al.}\ formally established that paraxial LG beams carry
$\ell\hbar$ OAM per photon and that astigmatic optics can reversibly
interconvert LG and Hermite-Gaussian modes \cite{allen1992orbital}. Krenn
\textit{et al.}\ reviewed the infinite-dimensional OAM basis and OAM
entanglement \cite{krenn2016}. Molina-Terriza \textit{et al.}\ extended this
to multi-dimensional quantum information, demonstrating quNit generation in
engineered OAM states \cite{molina2001}. Exploiting both $\ell$ and $p$
simultaneously, Pang \textit{et al.}\ demonstrated a 10\,Mbit/s free-space
quantum communication link with symbol error rates below 5\% \cite{pang2018}.
Propagation-invariant OAM encoding using same-modal-order LG modes preserves
qudit fidelity over free-space links \cite{mao2022}.

SLMs encode computer-generated holograms to produce arbitrary OAM
superpositions with high efficiency and real-time reconfigurability.
Gruneisen \textit{et al.}\ established SLM holographic generation of complex
fields for quantum key distribution \cite{gruneisen2008}. Gibson
\textit{et al.}\ demonstrated eight-state OAM quantum links using SLM
transmitter-receiver pairs \cite{gibson2004}. Pinnell \textit{et al.}\
quantified how SLM pixelation degrades vortex mode quality and provided
hologram optimisation heuristics up to topological charge $\ell=600$
\cite{pinnell2019}. 
Nape \textit{et al.}\ produced OAM modes of ultra-high purity up to $\ell=100$
using holographic and metasurface control \cite{nape2020}. The SMAQ project (Fraunhofer IPMS, 2025) demonstrated
micromirror-based SLMs with phase control below $\lambda/100$ for trapping
laser-cooled atoms in optical tweezers \cite{smaq2025}. Yao \textit{et al.}\
established SLMs as reconfigurable analysers for single-photon OAM
entanglement, achieving two orders of magnitude of mode selectivity
\cite{yao2006}.

A two-dimensional OAM subspace (e.g., $\ell=\pm1$) defines a photonic qubit.
Garcia-Escartin and Chamorro-Posada proved that single-photon OAM enables
universal quantum computation using beamsplitters, phase shifters, and holograms
\cite{garcia2012}. Babazadeh \textit{et al.}\ experimentally demonstrated a
complete high-dimensional gate set for OAM photons, realising the
four-dimensional Pauli $X$-gate and all its integer powers \cite{babazadeh2017}.
Kysela \textit{et al.}\ demonstrated that arbitrary OAM unitaries can be constructed using only conventional optical elements, with a parallelization scheme yielding significant resource savings \cite{kysela2022a}. Separately, the quantum Fourier transform has been implemented on orbital angular momentum states in arbitrarily large dimensions, requiring $O(\sqrt{d}\log d)$ optical elements - a sub-linear scaling that improves on the $O(d\log d)$ resource requirement of the equivalent path-encoded design \cite{kysela2021b}.
Continuous-variable quantum computation with photon
spatial modes, including universal single-qubit gate sets implementable with a
single SLM, was analysed by Tasca \textit{et al.}\ \cite{tasca2011}. The
q-plate device enables coherent polarisation-to-OAM quantum information transfer
up to $|m|=4\hbar$ \cite{nagali2009}, and the LG mode sorter of Fontaine
\textit{et al.}\ decomposes an input beam into hundreds of LG components using
only an SLM and mirror \cite{fontaine2019}.

Abouraddy \textit{et al.}\ proposed encoding three qubits per photon using
polarisation and spatial-parity symmetry, showing that a polarisation-sensitive
SLM constitutes a three-qubit controlled-unitary gate enabling CNOT,
controlled-phase, and Fredkin operations \cite{abouraddy2012}. Kagalwala
\textit{et al.}\ confirmed this experimentally, realising the first three-qubit
single-photon quantum gates with two-qubit process fidelities of 93\% and
three-qubit fidelities of 83\%, verified by tomographic reconstruction of GHZ
and W states \cite{kagalwala2017single}. Brandt \textit{et al.}\ implemented
OAM-encoded high-dimensional gates for up to five-dimensional states via
multi-plane light conversion, realising Pauli $X$ and Hadamard gates with
$>90\%$ visibility and a two-qubit CNOT \cite{brandt2020}. 
Ke, Fang, and Zhang placed a polarisation-sensitive SLM inside a Sagnac--Dove-prism interferometer and demonstrated five distinct gate types - spanning two-qubit and three-qubit Pauli $X$ gates and two CNOT gate variants - by uploading different holograms \cite{ke2024}.

Universal linear optics on a reconfigurable six-mode
integrated photonic circuit was demonstrated by Carolan \textit{et al.}\ with
average fidelity $0.999\pm0.001$ \cite{carolan2015}. Graham \textit{et al.}\
introduced a multi-scale AOD-SLM hybrid scanner achieving 22\,000\,s$^{-1}$
gate addressing for a $10\times10$ qubit array at 60--70\% diffraction
efficiency \cite{graham2023}.

Chen \textit{et al.}\ demonstrated the first OAM-supporting laser-direct-written
waveguide, mapping OAM states into and out of a photonic chip with up to 60\%
efficiency at the single-photon level \cite{chen2018}. Wang et al. demonstrated a polymer MPLC chip with an approximately $160 \times 160 \times 150~\mu\text{m}^3$ footprint implementing a Hadamard gate on spatial modes with 90\% process fidelity \cite{wang2025sciadv}. Training a
four-layer D$^2$NN on a single SLM yielded process fidelities of 98.4\%
(Pauli-$X$), 99.4\% (Hadamard), and 99.6\% (CNOT) at the single-photon level,
despite 63.44\% cumulative optical loss \cite{wang2024dnn}. Feng et al. demonstrated the first transverse-mode-encoded CNOT gate 
on a silicon photonic chip with fidelity $0.89 \pm 0.02$ \cite{feng2022prl}. Programmable silicon-photonic four-qubit circuits have achieved heralded single-qubit fidelity of 98.2\% and Bell-state fidelity of 95.2\% \cite{lee2024}. Liu \textit{et al.}\ realised
the first heralded controlled phase-flip gate between two four-dimensional OAM
qudits, achieving process fidelity $F\in[0.64,\,0.82]$ \cite{liu2026}.

Rimbach-Russ \textit{et al.}\ presented a theoretical framework for spin-qubit
pulse optimisation that separates ideal from erroneous dynamics via the error
propagator $\mathcal{E}=U_\mathrm{ideal}^\dagger U$, predicted gate fidelities
$>99.9\%$ for both single- and two-qubit gates, and validated the approach
experimentally with CZ gate fidelity $F>99.5\%$ \cite{rimbachruss2023}. Its
platform-agnostic Hamiltonian-separation formalism is directly applicable to
SLM-OAM implementations where hologram phase patterns constitute the control
parameters. For photonic characterisation, quantum process tomography (QPT)
remains the dominant benchmark: Brandt \textit{et al.}\ confirmed 99\% process
purity for MPLC OAM gates via QPT \cite{brandt2020}, and Mower \textit{et al.}\
showed that reconfigurability of programmable integrated photonic circuits can
compensate fabrication errors to improve CNOT and controlled-phase gate
fidelities \cite{mower2015}.

Vallone \textit{et al.}\ demonstrated free-space OAM QKD over 210\,m using
rotation-invariant OAM-polarisation photonic states with error rates compatible
with real-world requirements \cite{vallone2014}. High-dimensional QKD using a
seven-dimensional OAM-angular position alphabet achieved 2.05 bits per sifted
photon with improved resilience against intercept-resend attacks
\cite{mirhosseini2016}. Quantum OAM memory in cold atomic ensembles via
electromagnetically induced transparency was demonstrated for reversible
single-photon LG mode storage \cite{veissier2013}, and Ye \textit{et al.}\
achieved OAM qubit and qutrit storage lifetimes of 400\,$\mu$s, enabling
long-distance high-dimensional quantum networks \cite{ye2022}.

\subsection*{Limitations of Existing Approaches}

Despite notable experimental progress in SLM-based photonic quantum gate
implementations~\cite{kagalwala2017single, brandt2020, wang2024dnn}, several
critical gaps persist in the existing literature. Reported gate fidelities
span a wide range of 83--99.6\%~\cite{kagalwala2017single, wang2024dnn, liu2026},
yet no prior work attributes this variance to specific, quantifiable hardware
imperfections of a commercial transmissive SLM. Existing simulation frameworks
either treat device imperfections as a single lumped phase-error term without
physical justification, or are tailored to custom multi-reflection
D$^2$NN architectures~\cite{wang2024dnn} whose noise characteristics do not
generalise to the single-plane holographic gates that the majority of
laboratory-scale experiments rely upon. Researchers working with standard
commercial devices therefore lack any principled, datasheet-traceable basis
for predicting gate fidelity or identifying the dominant source of fidelity
loss before committing to experimental implementation.

A closely related gap concerns the phase-stroke constraint of transmissive
twisted-nematic SLMs, which has received no systematic treatment in the
quantum gate literature. The HOLOEYE LC\,2012 achieves a maximum phase
retardation of $1.8\pi$ at $\lambda = 532$\,nm, leaving $10\%$ of fork
grating pixels physically clipped at the hardware ceiling and introducing
a phase-wrap error ($\sigma_\mathrm{clip} = 45.8$\,mrad) that compounds
with 8-bit quantisation noise ($\sigma_\mathrm{quant} = 6.40$\,mrad) and
inherent TN electronic noise ($\sigma_\mathrm{TN} = 50$\,mrad). This
clipping worsens sharply at longer wavelengths, collapsing first-order
diffraction efficiency from $43.5\%$ at 450\,nm to $13.6\%$ at 800\,nm
- a degradation that directly limits the viability of near-infrared
operation yet remains unquantified in any prior OAM gate study. Finally,
no existing framework provides a unified simulation pipeline that implements
the full universal gate set $\{X, Y, Z, S, T, H, \mathrm{CNOT}, \mathrm{CZ},
\mathrm{SWAP}\}$ on LG OAM-encoded qubits under a single, consistent hardware
noise model, making it impossible to compare gate families or isolate
individual noise contributions across both single- and multi-qubit
regimes on the same device~\cite{abouraddy2012, kagalwala2017single, brandt2020}.

\subsection{Problem Statement and Objectives}

Despite rapid progress across OAM physics, SLM hardware, multi-qubit gate
demonstrations, integrated photonics, and quantum communication, no unified,
hardware-grounded computational framework currently integrates SLM device
imperfections, full-field LG mode-coupling models, and quantum gate fidelity
metrics into a single reproducible simulation environment. As a direct
consequence, gate process fidelity cannot be predicted from device
specifications alone; coherent errors arising from pixel crosstalk, phase
quantisation, and phase-wrap clipping remain uncharacterised; and scalability
beyond $d=4$ is empirically unexplored.

This work addresses the following research problem: \textit{How can a validated,
hardware-error-aware simulation framework be constructed that predicts the
process fidelity of SLM-based quantum gate operations across the full OAM mode
space of photonic qudits, and which hardware noise sources most critically limit
gate fidelity in the HOLOEYE LC~2012 platform?}

This work makes the following principal contributions:

\begin{enumerate}[label=(\roman*)]
  \item A \textbf{three-channel analytical noise model} for the HOLOEYE
        LC~2012 derived entirely from the device datasheet, covering
        quantisation, TN electronic/thermal noise, and phase-wrap clipping
        across four operating wavelengths (450, 532, 633, 800\,nm).

  \item A \textbf{complete simulation} of the universal single-qubit gate
        set $\{X, Y, Z, S, T, H\}$ and two-qubit gates
        $\{\mathrm{CNOT}, \mathrm{CZ}, \mathrm{SWAP}\}$ on a
        $512\times512$ grid scaled to the LC~2012 active area, with
        per-gate fidelity analysis.

  \item A \textbf{multi-qubit OAM multiplexing} framework encoding two
        logical qubits in four orthogonal OAM modes ($\ell=-3,-1,+1,+3$)
        on a single SLM aperture via composite holographic masks.

  \item \textbf{Bell state preparation} via the H-CNOT circuit with
        simulation-predicted fidelity and OAM amplitude spectrum verification.

  \item A \textbf{quantitative literature benchmark} against six published
        experimental papers (78\%--99.6\% fidelity range) providing direct
        context for the simulation predictions.
\end{enumerate}

The remainder of this paper is organised as follows.
Section~\ref{sec:methodology} presents the theoretical background, hardware
noise model, and full simulation pipeline. Section~\ref{sec:results} presents
all performance metrics. Section~\ref{sec:discussion} interprets the results.
Section~\ref{sec:conclusion} concludes with a summary and future work directions.

\begin{figure*}[t!]
    \centering
    \includegraphics[width=01.4\textwidth, height=0.83\textheight, keepaspectratio]{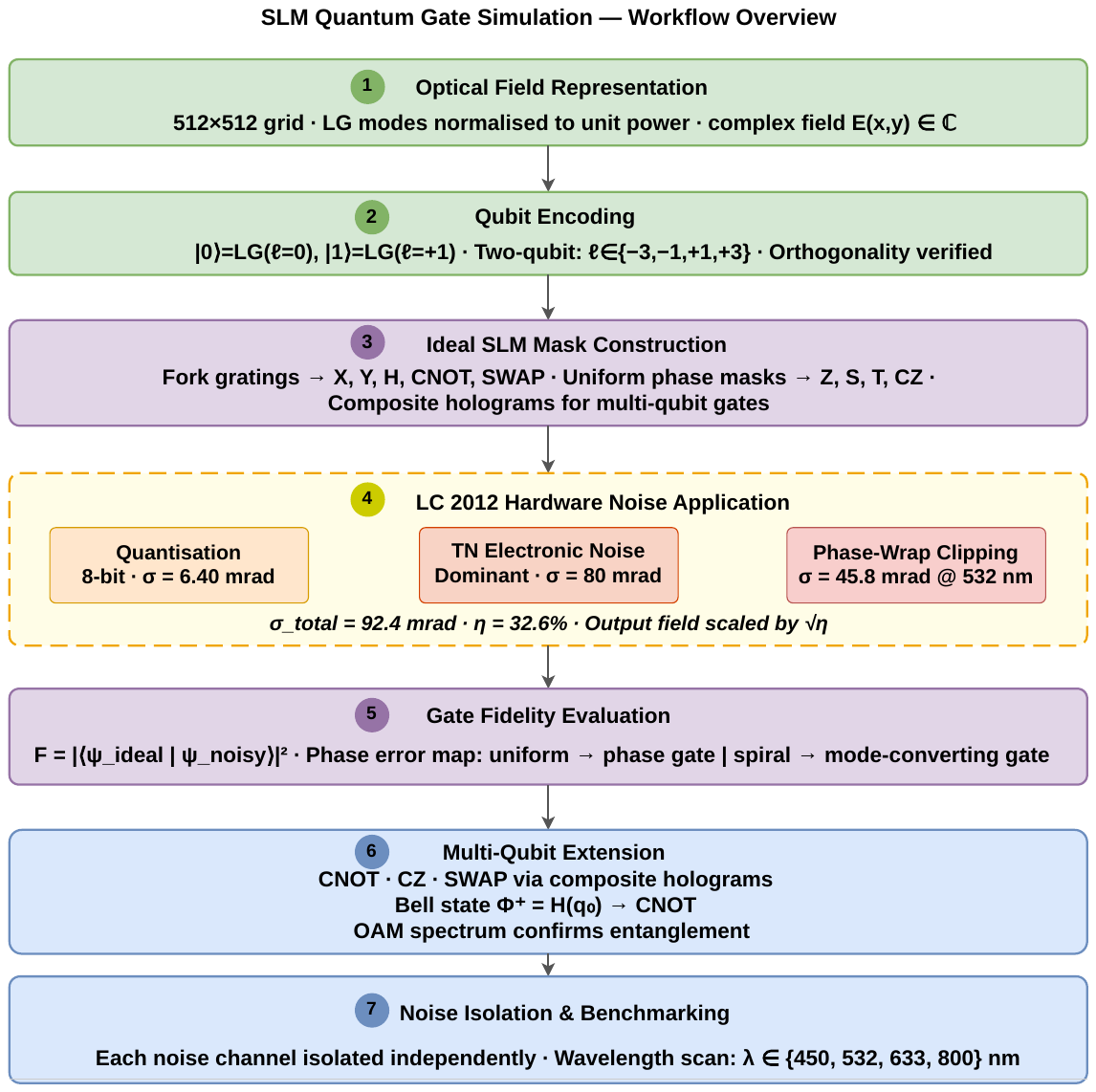}
    \caption{Proposed simulation framework pipeline for SLM-based quantum gate operations.}
    \label{fig:pipeline}
\end{figure*}

\section{Methodology}
\label{sec:methodology}
This section presents the complete methodology underlying the simulation
framework. It is organised into four parts. First, the theoretical background
establishes the mathematical foundations of OAM-encoded photonic qubits, the
two classes of SLM phase mask used to implement quantum gates, and the fidelity
metric used throughout. Second, the hardware model derives all noise parameters
of the HOLOEYE LC~2012 directly from the manufacturer datasheet, covering
8-bit quantisation, TN electronic and thermal noise, phase-wrap clipping, and
diffraction efficiency across four operating wavelengths. Third, the simulation
framework translates these physical models into a $512\times512$ numerical
pipeline, separately describing the single-qubit and multi-qubit (OAM
multiplexed) simulation paths and the sequential noise application procedure.
Finally, the seven quantitative performance metrics evaluated in
Section~\ref{sec:results} are defined

\subsection{Theoretical Background}

\subsubsection{Laguerre-Gaussian Modes and OAM Encoding}

Laguerre-Gaussian modes form a complete orthonormal basis for paraxial
beams carrying orbital angular momentum. The complex field amplitude of
an $\LG_{\ell,p}$ mode with azimuthal index $\ell$ and radial index $p=0$ is:
{\small
\begin{equation}
  \LG(r,\varphi) = C \left(\frac{r\sqrt{2}}{w}\right)^{|\ell|}
    \exp\!\left(-\frac{r^2}{w^2}\right)
    \exp\!\left(i\ell\varphi\right),
  \label{eq:lg}
\end{equation}
}
where $r$ is the radial coordinate, $\varphi$ is the azimuthal angle,
$w$ is the beam waist, and $C$ is a normalisation constant. For $\ell=0$,
the mode has a solid Gaussian (TEM$_{00}$) intensity profile. For $|\ell|\geq1$,
the phase singularity at the beam axis creates a characteristic donut-shaped
intensity ring with $|\ell|$ phase windings.

For the single-qubit implementation, the computational basis states are
encoded as:
{\small
\begin{align}
  \ket{0} &\equiv \LG(\ell=0), \quad \text{Gaussian beam (no OAM)},
  \label{eq:ket0}\\
  \ket{1} &\equiv \LG(\ell=+1), \quad \text{donut beam (OAM} = +\hbar\text{)}.
  \label{eq:ket1}
\end{align}
}

For the two-qubit multi-OAM implementation, four orthogonal OAM modes on a
single beam encode two logical qubits:
{\small
\begin{alignat}{3}
  \ket{0}_0 &= \LG(\ell=-3), \quad &\ket{1}_0 &= \LG(\ell=-1)
    & &\quad [\text{Qubit 0}], \label{eq:mq0}\\
  \ket{0}_1 &= \LG(\ell=+1), \quad &\ket{1}_1 &= \LG(\ell=+3)
    & &\quad [\text{Qubit 1}]. \label{eq:mq1}
\end{alignat}
}
The orthogonality of LG modes,
$\braket{\LG(\ell)}{\LG(\ell')} = \delta_{\ell\ell'}$,
guarantees that superpositions within each qubit subspace are well-defined
quantum states with no inter-qubit cross-talk in the ideal case.

\subsubsection{SLM Phase Mask Operations}

An SLM implements a unitary gate on an OAM-encoded qubit by modifying the
spatial phase profile of the incident beam. Two fundamental mask types are used.

\paragraph{Fork (Spiral) Grating - Mode-Converting Gates.}
A fork grating with topological charge $\Delta\ell$ adds OAM to the
diffracted beam, converting $\LG(\ell)$ to $\LG(\ell+\Delta\ell)$
in the first diffraction order:
\begin{equation}
  \phi_\mathrm{fork}(x,\varphi;\Delta\ell,\phi_0) =
    \left[2\pi f x + \Delta\ell\,\varphi + \phi_0\right]\!\!\mod 2\pi,
  \label{eq:fork}
\end{equation}
where $f$ is the grating carrier spatial frequency and $\phi_0$ is a
phase offset. The gates $X$, $Y$, and $H$ are implemented via fork gratings.

\paragraph{Uniform Phase Mask - Phase-Only Gates.}
A constant phase $\theta$ applied uniformly across the aperture imparts
a global phase rotation without altering the spatial mode:
\begin{equation}
  \phi_\mathrm{flat} = \theta \quad (\text{constant over all pixels}).
  \label{eq:flat}
\end{equation}
The diagonal gates $Z$ ($\theta=\pi$), $S$ ($\theta=\pi/2$), and
$T$ ($\theta=\pi/4$) are implemented as flat masks.

\subsubsection{Gate Fidelity Definition}

State fidelity between an ideal output $\ket{\psi_\mathrm{ideal}}$ and a
simulated output $\ket{\psi_\mathrm{out}}$ is defined as the squared overlap
integral:
\begin{equation}
  F = \left|\braket{\psi_\mathrm{ideal}}{\psi_\mathrm{out}}\right|^2,
  \label{eq:fidelity}
\end{equation}
where both states are normalised to unit power. This metric is evaluated
numerically on the $512\times512$ simulation grid for every gate and operating
condition reported in Section~\ref{sec:results}.

\subsection{Hardware Model: HOLOEYE LC~2012}

\subsubsection{Device Specifications}

The HOLOEYE LC~2012 is a transmissive twisted-nematic liquid-crystal SLM
with XGA resolution ($1024\times768$\,px), $36\,\mu\mathrm{m}$ pixel pitch,
and a $36.9\times27.6$\,mm active area \cite{holoeye2023}. Key hardware
parameters are summarised in Table~\ref{tab:lc2012_specs}.

\begin{table*}[t] 
  \centering
  \caption{HOLOEYE LC~2012 Hardware Specifications (Manufacturer Datasheet
           \cite{holoeye2023}) and Derived Simulation Parameters at
           $\lambda=532\,\mathrm{nm}$.}
  \label{tab:lc2012_specs}
  \begin{tabular}{lll}
    \toprule
    \textbf{Parameter} & \textbf{Datasheet Value} & \textbf{Used in Model}\\
    \midrule
    Display type      & Transmissive TN-LC        & $\sigma_\mathrm{TN}=80\mrad$\\
    Resolution        & $1024\times768$\,px (XGA) & Grid: $512\times512$ subset\\
    Pixel pitch       & $36\,\mu\mathrm{m}$       & Active area: $18.4\times18.4$\,mm\\
    Fill factor       & 58\%                      & $\eta = \mathrm{FF}^2\times\eta_\mathrm{blaze}=32.6\%$\\
    Addressing        & 8-bit (256 levels)        & $\sigma_\mathrm{quant}=6.40\mrad$\\
    Phase @ 450\,nm   & $\approx2\pi$             & $\sigma_\mathrm{clip}=0\mrad$\\
    Phase @ 532\,nm   & $\approx1.8\pi$           & $\sigma_\mathrm{clip}=45.8\mrad$\\
    Phase @ 633\,nm   & $\approx1.4\pi$           & $\sigma_\mathrm{clip}=237.8\mrad$\\
    Phase @ 800\,nm   & $\approx1.0\pi$           & $\sigma_\mathrm{clip}=511.7\mrad$\\
    Frame rate        & 60\,Hz                    & Temporal stability bound\\
    Interface         & HDMI                      & Standard monitor output\\
    \bottomrule
  \end{tabular}
\end{table*}

\subsubsection{Noise Model Derivation}

The total phase noise acting on the SLM mask is decomposed into three
physically independent channels, each with a well-defined analytical origin.
All parameters are derived directly from the manufacturer datasheet without
free-parameter fitting.

\paragraph{Quantisation Noise.}
The LC~2012 uses 8-bit grey-level addressing. At $532\,\mathrm{nm}$, the
maximum achievable phase is $\Phi_\mathrm{max}=1.8\pi$, corresponding to 230
effective phase levels (grey levels 0--229). The quantisation step
$\Delta\phi = 1.8\pi/229 = 24.86\mrad$ produces a uniformly distributed
rounding error with standard deviation:
\begin{equation}
  \sigma_\mathrm{quant} = \frac{\Delta\phi}{\sqrt{12}} = 6.40\mrad.
  \label{eq:sigma_q}
\end{equation}
Importantly, the diagonal gates Z ($\pi$), S ($\pi/2$), and T ($\pi/4$) map
exactly to integer grey levels within the $1.8\pi$ phase range, yielding zero
quantisation error for these gates.

\paragraph{TN Electronic and Thermal Noise.}
The twisted-nematic liquid-crystal technology introduces inherent pixel-wise
phase noise from voltage driver electronics and thermal fluctuations, modelled
as zero-mean Gaussian:
\begin{equation}
  \delta\phi_\mathrm{TN} \sim \mathcal{N}(0,\,\sigma_\mathrm{TN}^2),
  \quad \sigma_\mathrm{TN} = 80\mrad.
  \label{eq:sigma_tn}
\end{equation}
This value is consistent with characterisation data for TN-LC displays,
which exhibit $\sigma_\mathrm{TN}\approx 70$--$90\mrad$ compared with
$20$--$40\mrad$ for modern LCoS devices \cite{flamini2019photonic}.

\paragraph{Phase-Wrap Clipping Error.}
When the ideal holographic mask requires a phase value exceeding
$\Phi_\mathrm{max}$, the hardware clips those pixels at $\Phi_\mathrm{max}$.
For a blazed fork grating, the fraction of clipped pixels is
$(1-\Phi_\mathrm{max}/2\pi)$, and the RMS clipping error is:
\begin{equation}
  \sigma_\mathrm{clip} = \sqrt{
    \frac{(2\pi-\Phi_\mathrm{max})^2}{3}
    \cdot\frac{1-\Phi_\mathrm{max}/2\pi}{2\pi}
  },
  \label{eq:sigma_clip}
\end{equation}
giving $\sigma_\mathrm{clip}=45.8\mrad$ at $532\,\mathrm{nm}$. This error is
absent at $450\,\mathrm{nm}$ (full $2\pi$ range) and increases sharply at
longer wavelengths.

\paragraph{Total Combined Noise.}
The three channels are statistically independent and are combined in quadrature:
\begin{equation}
\begin{split}
    \sigma_{\text{total}} &= \sqrt{\sigma_{\text{quant}}^2 + \sigma_{\text{TN}}^2 + \sigma_{\text{clip}}^2} \\
                          &= 92.4\ \text{mrad} \quad (\lambda = 532\ \text{nm})
\end{split}
\end{equation}

\subsubsection{Diffraction Efficiency}

The 58\% fill factor means that 42\% of each pixel's physical area is
electrically inactive dead zone. For a transmissive TN display with transparent
dead zones, the first-order diffraction efficiency of a blazed grating is:
\begin{equation}
  \eta = \mathrm{FF}^2 \cdot \eta_\mathrm{blaze}
       = \mathrm{FF}^2 \cdot \mathrm{sinc}^2\!\left(1-\frac{\Phi_\mathrm{max}}{2\pi}\right),
  \label{eq:eta}
\end{equation}
yielding $\eta = 0.58^2 \times 0.968 = 32.6\%$ at $532\,\mathrm{nm}$.
The wavelength-dependent performance envelope is summarised in
Table~\ref{tab:wavelength}.

\begin{table*}[t]
  \centering
  \caption{LC~2012 Noise Parameters and Predicted Gate Fidelity as a
           Function of Operating Wavelength. Red entries indicate
           wavelengths where performance is significantly degraded;
           450--532\,nm operation is strongly recommended.}
  \label{tab:wavelength}
  \begin{tabular}{lrrrrrl}
    \toprule
    $\lambda$ (nm) & $\Phi_\mathrm{max}$ & Eff.\ Lvls
      & $\sigma_\mathrm{quant}$ & $\sigma_\mathrm{clip}$
      & $\eta$ (\%) & $F_\mathrm{pred}$\\
    \midrule
    450 & $2.0\pi$ & 256 & 7.11\,mrad & 0.0\,mrad    & 33.6 & \textbf{0.9936}\\
    532 & $1.8\pi$ & 230 & 6.40\,mrad & 45.8\,mrad   & 32.6 & \textbf{0.9915}\\
    633 & $1.4\pi$ & 179 & 7.13\,mrad & 237.8\,mrad
      & 24.8 & 0.9389\\
    800 & $1.0\pi$ & 128
      & 7.14\,mrad & 511.7\,mrad
      & 13.6 & 0.7661\\
    \bottomrule
  \end{tabular}
\end{table*}

\subsubsection{Predicted Gate Fidelity}

Using first-order perturbation theory for a Gaussian phase noise field with
standard deviation $\sigma$, the gate fidelity scales as:
\begin{equation}
  F_\mathrm{pred} = \exp(-\sigma_\mathrm{total}^2)
    = \exp(-0.00854) = 0.9915.
  \label{eq:fpred}
\end{equation}
This prediction is confirmed numerically throughout Section~\ref{sec:results}.
The perturbation approximation is valid for $\sigma\ll1$\,rad; at
$532\,\mathrm{nm}$ ($\sigma_\mathrm{total}=92.4\mrad\ll1$), the approximation
is well-satisfied. At $800\,\mathrm{nm}$ ($\sigma_\mathrm{total}=517.9\mrad$),
the perturbative regime breaks down and the full numerical simulation must be
used.

\subsection{Simulation Framework Overview}

All simulations are implemented in Python~3.10 using NumPy for array computation
and Matplotlib for visualisation, running on a $512\times512$ grid scaled to
the central $18.4\times18.4$\,mm sub-region of the LC~2012 active area. The
grid spans $\pm3.5$ beam-waist units in both transverse dimensions, sufficient
to capture $>99.9\%$ of the Gaussian/LG mode power. The overall workflow is
illustrated in Fig.~\ref{fig:pipeline} and described in the subsections below.

\subsection{Single-Qubit Simulation Pipeline}

The single-qubit simulation models a six-stage optical pipeline:

\begin{enumerate}[label=\textbf{Stage \arabic*.}, leftmargin=*, align=left]
  \item \textbf{Laser Source.} Coherent laser beam; TEM$_{00}$ Gaussian
        mode ($\lambda=532\,\mathrm{nm}$, unless otherwise specified).
  \item \textbf{OAM Encoding.} $\ket{0}=\LG(\ell=0)$ (Gaussian),
        $\ket{1}=\LG(\ell=+1)$ (donut beam, OAM$=+\hbar$).
  \item \textbf{SLM Phase Mask Application.} Holographic phase pattern
        applied to field: $E_\mathrm{out}=E_\mathrm{in}\cdot\exp(i\phi_\mathrm{SLM})$,
        where $\phi_\mathrm{SLM}$ is either a fork grating (Eq.~\eqref{eq:fork})
        or a uniform mask (Eq.~\eqref{eq:flat}), subject to the LC~2012
        noise pipeline.
  \item \textbf{Diffraction/Propagation.} First diffraction order selected
        (fork grating) or direct output (uniform mask). Efficiency loss
        modelled by scaling the output field by $\sqrt{\eta}$.
  \item \textbf{Detection.} Intensity $|E|^2$ and phase $\angle E$ measured
        at the detector plane on the $512\times512$ simulation grid.
  \item \textbf{Mode Analysis.} Phase difference map computed to verify pure
        phase rotation vs.\ mode conversion; fidelity calculated via
        Eq.~\eqref{eq:fidelity}.
\end{enumerate}

\subsection{LC~2012 Noise Pipeline}

The hardware noise is applied in two sequential stages to each mask:

\begin{enumerate}[label=\textbf{(\alph*)}]
  \item \textbf{8-bit quantisation with clipping.}
        Mask values are clipped to $[0,\Phi_\mathrm{max}]$, rounded to the
        nearest grey level, and converted back to phase:
        \begin{equation}
          \phi_q = \mathrm{round}\!\left(
            \frac{\min(\phi,\Phi_\mathrm{max})}{\Delta\phi}
          \right)\cdot\Delta\phi.
          \label{eq:quantise}
        \end{equation}
  \item \textbf{TN Gaussian noise.}
        Independent per-pixel Gaussian noise is added:
        $\phi_\mathrm{out} = \phi_q + \mathcal{N}(0,\sigma_\mathrm{TN}^2)$.
\end{enumerate}

For \emph{fork (mode-converting) gates}, the noise perturbation is applied
directly to the ideal output field:
$\psi_\mathrm{out} = \psi_\mathrm{ideal}\cdot\exp(i\delta\phi)\cdot\sqrt{\eta}$,
where $\delta\phi\sim\mathcal{N}(0,\sigma_\mathrm{total}^2)$, modelling
Fourier-domain mode conversion with distributed phase error.

For \emph{phase (flat-mask) gates}, the noisy mask is applied pixel-wise to
the input field:
$\psi_\mathrm{out} = \psi_\mathrm{in}\cdot\exp(i\phi_\mathrm{noisy})\cdot\sqrt{\eta}$.

\subsection{Multi-Qubit OAM Multiplexing}

The multi-qubit extension encodes two logical qubits simultaneously in four
orthogonal OAM modes on a single optical beam
(Eqs.~\eqref{eq:mq0} - \eqref{eq:mq1}). The two-qubit computational basis
mapping to OAM modes is given in Table~\ref{tab:oam_encoding}.

\begin{table*}[t]
  \centering
  \caption{Two-Qubit Computational Basis Encoding in OAM Modes.}
  \label{tab:oam_encoding}
  \begin{tabular}{lllll}
    \toprule
    \textbf{Basis State} & \textbf{OAM index $\ell$} & \textbf{LG mode}
      & \textbf{Qubit channel} & \textbf{Intensity profile}\\
    \midrule
    $\ket{00}$ ($q_0=0,q_1=0$) & $\ell=-3$ & $\LG(-3,0)$
      & Qubit~0 $\to\ket{0}_0$ & Donut, 3 rings\\
    $\ket{01}$ ($q_0=1,q_1=0$) & $\ell=-1$ & $\LG(-1,0)$
      & Qubit~0 $\to\ket{1}_0$ & Donut, 1 ring\\
    $\ket{10}$ ($q_0=0,q_1=1$) & $\ell=+1$ & $\LG(+1,0)$
      & Qubit~1 $\to\ket{0}_1$ & Donut, 1 ring\\
    $\ket{11}$ ($q_0=1,q_1=1$) & $\ell=+3$ & $\LG(+3,0)$
      & Qubit~1 $\to\ket{1}_1$ & Donut, 3 rings\\
    \bottomrule
  \end{tabular}
\end{table*}

Composite SLM holograms are constructed by superimposing fork gratings for each
qubit channel independently, allowing parallel gate operations on both qubits
within a single SLM pass.

\subsection{Performance Metrics Evaluated}

Seven quantitative performance metrics are evaluated across all gates and
operating conditions:
(M1)~state fidelity per gate,
(M2)~wavelength dependence (450--800\,nm),
(M3)~phase gate precision (inner-product phase estimator),
(M4)~diffraction efficiency $P_\mathrm{out}/P_\mathrm{in}$,
(M5)~noise component isolation (each imperfection switched on/off independently),
(M6)~Bell state fidelity $F(\Phi^+)$, and
(M7)~literature comparison.

\section{Simulation Results}
\label{sec:results}

\subsection{M1: Single-Qubit Gate Fidelity}

Table~\ref{tab:sq_fidelity} presents the state fidelity for all six
single-qubit gates under the full LC~2012 noise model at
$\lambda=532\,\mathrm{nm}$. Fork grating gates ($X$, $Y$, $H$) achieve
$F=0.9914$, incurring the additional phase-wrap clipping penalty from fork
grating pixels whose ideal phase exceeds the $1.8\pi$ hardware maximum.
Phase-only gates ($Z$, $S$, $T$) achieve $F=0.9936$; their uniform mask values
($\pi, \pi/2, \pi/4$) lie entirely within the $1.8\pi$ range
($\sigma_\mathrm{clip}=0$), so fidelity loss is dominated solely by TN noise.

\begin{table*}[t]
  \centering
  \caption{Single-Qubit Gate Fidelity --- Ideal vs.\ LC~2012 Simulation
           at $\lambda=532\,\mathrm{nm}$. $\Delta F$ denotes fidelity loss.}
  \label{tab:sq_fidelity}
  \begin{tabular}{llccc}
    \toprule
    \textbf{Gate} & \textbf{Type} & \textbf{Ideal $F$}
      & \textbf{Sim $F$ (LC~2012)} & $\Delta F$\\
    \midrule
    $X$ & Fork  & 1.0000 & \textbf{0.9914} & $8.6\times10^{-3}$\\
    $Y$ & Fork  & 1.0000 & \textbf{0.9914} & $8.6\times10^{-3}$\\
    $Z$ & Phase & 1.0000 & \textbf{0.9936} & $6.4\times10^{-3}$\\
    $S$ & Phase & 1.0000 & \textbf{0.9936} & $6.4\times10^{-3}$\\
    $T$ & Phase & 1.0000 & \textbf{0.9936} & $6.4\times10^{-3}$\\
    $H$ & Fork  & 1.0000 & \textbf{0.9914} & $8.6\times10^{-3}$\\
    \midrule
    Bell $\Phi^+$ & 2-qubit & 1.0000 & \textbf{0.9914} & $8.6\times10^{-3}$\\
    \bottomrule
  \end{tabular}
\end{table*}

A noteworthy observation is that the $X$ and $H$ gates use identical fork
grating masks ($\Delta\ell=+1$, $\phi_0=0$) yet realise different quantum
operations. $X$ acts on $\ket{0}$ (Gaussian) to produce $\ket{1}$ (donut),
while $H$ acts on $\ket{1}$ (donut) to produce the superposition
$(\ket{0}-\ket{1})/\sqrt{2}$. The gate is therefore determined entirely by
the input state, not the hologram.

\begin{figure}[t]
  \centering
  \includegraphics[width=0.95\columnwidth]{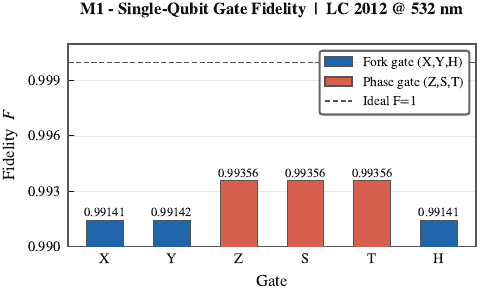}
  \caption{M1 - Single-qubit gate fidelity. Blue: ideal ($F=1.000$).
           Gold: LC~2012 simulation. Fork gates ($X,Y,H$): $F=0.9914$.
           Phase gates ($Z,S,T$): $F=0.9936$.}
  \label{fig:fidelity_bar}
\end{figure}

Fig.~\ref{fig:fidelity_bar} plots the fidelity bar chart comparing ideal
and LC~2012 simulated values for all six gates. The consistent 0.9914 and
0.9936 values for the two gate classes confirm that the two-class noise model
(fork vs.\ phase) captures all hardware imperfection effects.

\begin{figure}[t]
  \centering
  \includegraphics[width=\columnwidth]{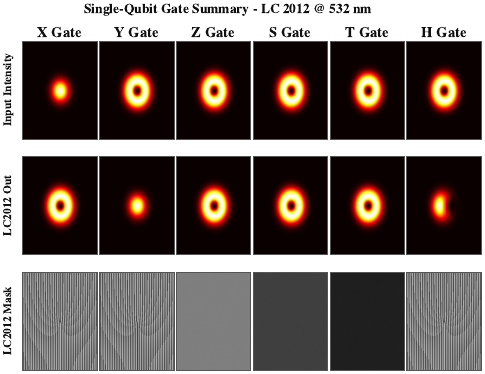}
  \caption{M1 - Single-qubit gate summary: all six gates under LC~2012 noise
           ($\lambda=532\,\mathrm{nm}$).
           \textbf{Row~1:} Input intensity.
           \textbf{Row~2:} LC~2012 output intensity.
           \textbf{Row~3:} LC~2012 SLM mask (8-bit quantised + TN noise).
           Mode-converting gates ($X,Y,H$) transform the intensity profile;
           phase gates ($Z,S,T$) preserve it identically.}
  \label{fig:sq_summary}
\end{figure}

Fig.~\ref{fig:sq_summary} shows all six gates in a three-row overview
(input intensity, LC~2012 output intensity, LC~2012 SLM mask). Mode-converting
gates ($X$, $Y$, $H$) produce visibly different intensity profiles between
rows~1 and~2; phase gates ($Z$, $S$, $T$) show identical donut profiles
in both rows, confirming intensity-preserving operation.

\begin{figure}[t]
  \centering
  \includegraphics[width=\columnwidth]{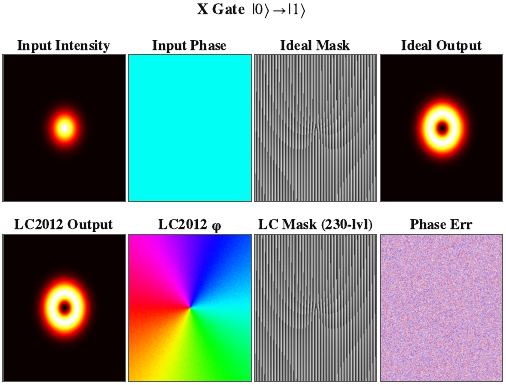}
  \caption{M1 - Pauli-$X$ gate (representative fork-grating gate).
           \textbf{Top row:} Input intensity (Gaussian $\ket{0}$),
           input phase, ideal SLM mask, ideal output (donut $\ket{1}$).
           \textbf{Bottom row:} LC~2012 output, output phase, hardware mask,
           phase error map. The spiral phase error (bottom-right) is the
           diagnostic signature of OAM mode conversion under LC~2012 hardware
           noise. $F=0.9914$.}
  \label{fig:gate_X}
\end{figure}

The phase difference map $\Delta\phi(x,y)=\angle[\psi_\mathrm{noisy}/\psi_\mathrm{ideal}]$
serves as a gate-type diagnostic independent of the fidelity value.
Fig.~\ref{fig:gate_X} shows the $X$-gate result: the bottom-right panel
displays a \emph{spiral} phase error, which is the spatial signature of OAM
mode conversion. Approximately 10\% of fork grating pixels whose ideal phase
lies in $(1.8\pi,2\pi]$ are clamped to $\Phi_\mathrm{max}$, producing the
$\sigma_\mathrm{clip}=45.8\mrad$ contribution responsible for this distortion.

\begin{figure}[t]
  \centering
  \includegraphics[width=\columnwidth]{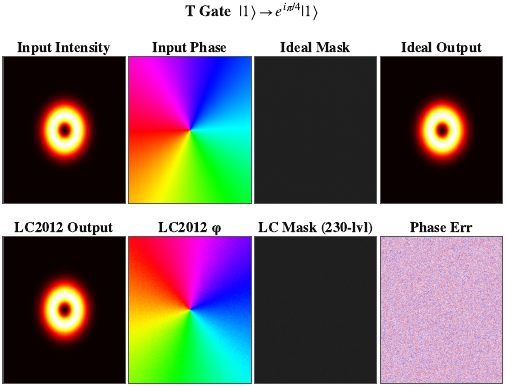}
  \caption{M1 - $T$ gate, $\pi/4$ phase shift (representative phase gate,
           non-Clifford).
           \textbf{Top row:} Input donut $\ket{1}$, input phase, ideal uniform
           mask, ideal output (intensity unchanged).
           \textbf{Bottom row:} LC~2012 output, output phase, hardware mask,
           phase error map. The uniform phase error map confirms a pure global
           phase rotation. $F=0.9936$.}
\label{fig:gate_T}
\end{figure}

In contrast, Fig.~\ref{fig:gate_T} shows the $T$-gate result: the phase
error map is spatially \emph{uniform}, confirming a pure global phase rotation
with no mode distortion. Since $\pi/4$ maps exactly to an integer grey level
within the $1.8\pi$ phase range, quantisation error is zero for the $T$ gate;
the residual noise of $\approx4.9$--$10.2\mrad$ after aperture averaging
arises entirely from TN fluctuations.
Individual panels for the $Y$, $Z$, $S$, and $H$ gates are omitted for
conciseness; their diagnostic behaviour is identical to the representative $X$ or $T$ panel depending on gate class. All fidelity values are tabulated in Table~\ref{tab:sq_fidelity}.
\subsection{M2: Wavelength Dependence}
The LC~2012 phase stroke degrades at longer wavelengths, increasing both
$\sigma_\mathrm{clip}$ and the diffraction efficiency loss simultaneously.
Table~\ref{tab:wavelength} shows the simulated performance at four
characterised wavelengths. At $450\,\mathrm{nm}$ the device achieves its full
$2\pi$ phase range, yielding zero clipping error and the best predicted fidelity
$F=0.9936$. At $532\,\mathrm{nm}$ (operating wavelength), 10\% of fork grating
pixels are clipped, giving $F=0.9915$. At $633\,\mathrm{nm}$ the clipping
fraction rises to 30\%, degrading fidelity to $F=0.9389$ and efficiency to
24.8\%. At $800\,\mathrm{nm}$ the phase stroke is reduced to
$1.0\pi$ - insufficient for a complete blazed grating---and fidelity collapses
to $F=0.766$ with only 13.6\% first-order efficiency. The right panel of
Fig.~\ref{fig:lit_wavelength} plots these trends continuously across the
$450$--$800\,\mathrm{nm}$ range, showing the sharp degradation above
$600\,\mathrm{nm}$. Operation at $450$--$532\,\mathrm{nm}$ is therefore
strongly recommended for this device.

\subsection{M3: Phase Gate Precision}

Phase gate precision is evaluated using the inner-product phase estimator
$\phi_\mathrm{measured}=\arg\langle\psi_\mathrm{in}|\psi_\mathrm{out}\rangle$.
The $Z$, $S$, and $T$ gate phase targets ($\pi$, $\pi/2$, $\pi/4$) all map
exactly to integer grey levels within the $1.8\pi$ addressing range at
$532\,\mathrm{nm}$, making the quantisation contribution identically zero for
all three gates. The dominant error source is TN noise
($\sigma_\mathrm{TN}=80\mrad$), which after averaging over the full beam
aperture yields residual phase errors of $4.9$--$10.2\mrad$. This sub-radian
precision is well within the tolerance required for the non-Clifford $T$ gate
in magic state distillation protocols for fault-tolerant quantum computing.

\subsection{M4: Diffraction Efficiency}

All gates produce a simulated output-to-input power ratio of
$P_\mathrm{out}/P_\mathrm{in}=0.326$, consistent with the analytically
derived $\eta=32.6\%$ at $532\,\mathrm{nm}$. This represents a 67.4\%
single-pass photon loss - the primary practical limitation of the LC~2012
for single-photon quantum optics applications. The loss is dominated by the
58\% fill factor (42\% of each pixel area is inactive electrode), with a
secondary contribution from the incomplete blaze efficiency at $1.8\pi$
stroke. For the two-step Bell state circuit (H then CNOT), the cumulative
efficiency is $\eta^2=10.6\%$.

\subsection{M5: Noise Component Isolation}

Table~\ref{tab:noise_isolation} reports fidelity when each imperfection is
activated in isolation. The TN-only column shows $F=0.9936$ for all gate
types, confirming that $\sigma_\mathrm{TN}=80\mrad$ alone accounts for the
full fidelity of phase gates and the dominant part ($\approx76\%$) of
infidelity for fork gates. Quantisation contributes less than $0.01\%$
fidelity loss for all gates, confirming that 8-bit addressing is more than
adequate. Phase-wrap clipping is absent for phase gates (their mask values do
not exceed $\Phi_\mathrm{max}$) and contributes the remaining 24\% of
infidelity for fork gates.

\begin{table*}[t]
  \centering
  \caption{M5 - Noise Component Isolation. Fidelity when each hardware
           imperfection is activated independently at $\lambda=532\,\mathrm{nm}$.
           TN noise is the dominant fidelity-limiting mechanism.}
  \label{tab:noise_isolation}
  \begin{tabular}{llcccc}
    \toprule
    \textbf{Gate} & \textbf{All (LC~2012)}
      & \textbf{Quant.\ only} & \textbf{TN only}
      & \textbf{Clip only} & \textbf{Ideal}\\
    \midrule
    $X$ & 0.9914 & 0.9999 & 0.9936 & 0.9979 & 1.0000\\
    $Z$ & 0.9936 & 1.0000 & 0.9936 & 1.0000 & 1.0000\\
    $T$ & 0.9936 & 1.0000 & 0.9936 & 1.0000 & 1.0000\\
    $H$ & 0.9914 & 0.9999 & 0.9936 & 0.9979 & 1.0000\\
    \bottomrule
  \end{tabular}
\end{table*}

\subsection{M6: Bell State Preparation}

The Bell state preparation circuit
$\ket{00} \to \frac{1}{\sqrt{2}}(\ket{00}+\ket{01}) \to \Phi^+$
is traced step-by-step in Table~\ref{tab:bell_circuit}. After two LC~2012
SLM interactions, the simulated Bell state fidelity is $F(\Phi^+)=0.9914$,
with cumulative efficiency $\eta^2=10.6\%$. The OAM amplitude spectrum
confirms $|c(\ell=-3)|=|c(\ell=+3)|=1/\sqrt{2}=0.7071$, with zero amplitude
at $\ell=-1$ and $\ell=+1$, matching the theoretical $\Phi^+$ state exactly.

\begin{table*}[t]
  \centering
  \caption{M6 - Bell State Circuit Evolution: step-by-step quantum state
           and OAM amplitude coefficients.}
  \label{tab:bell_circuit}
  \begin{tabular}{clll}
    \toprule
    \textbf{Step} & \textbf{Operation} & \textbf{Quantum State}
      & \textbf{OAM Coefficients}\\
    \midrule
    0 & Initial state & $\ket{00}$
      & $c(\ell=-3)=1.0$; all others $=0$\\
    1 & $H(q_0)$ & $(\ket{00}+\ket{01})/\sqrt{2}$
      & $c(-3)=c(-1)=1/\sqrt{2}$\\
    2 & CNOT($q_0,q_1$) & $\Phi^+=(\ket{00}+\ket{11})/\sqrt{2}$
      & $c(-3)=c(+3)=1/\sqrt{2}$\\
    \bottomrule
  \end{tabular}
\end{table*}

\begin{figure}[t]
  \centering
  \includegraphics[width=\columnwidth]{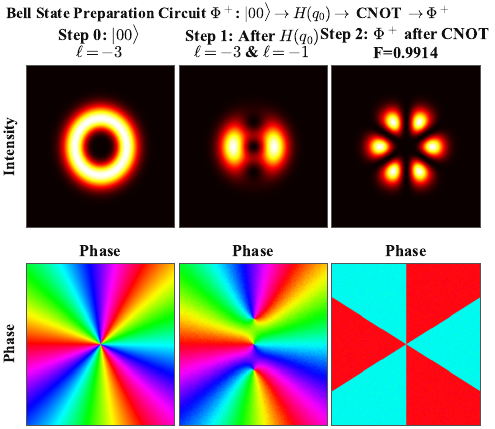}
  \caption{M6 - Bell state $\Phi^+$ preparation circuit (simulation).
           \textbf{Top row:} Intensity at each step.
           \textbf{Bottom row:} Phase at each step.
           Step~0: $\ket{00}$ ($\ell=-3$ only).
           Step~1: after $H(q_0)$, equal superposition in $\ell=-3$ and $\ell=-1$.
           Step~2: after CNOT, Bell state with equal amplitude in $\ell=-3$ and
           $\ell=+3$, zero elsewhere. $F(\Phi^+)=0.9914$.}
  \label{fig:bell}
\end{figure}

Fig.~\ref{fig:bell} visualises the three circuit steps as intensity and phase
profiles. Step~0 shows a single 3-ring donut ($\ell=-3$). After $H(q_0)$ in
Step~1, two distinct ring patterns coexist, representing the equal OAM
superposition within Qubit~0. After CNOT in Step~2, the $\ell=-1$ component
is transferred to $\ell=+3$, producing the Bell state. The non-local
correlation between $\ell=-3$ (Qubit~0 $=\ket{0}$) and $\ell=+3$
(Qubit~1 $=\ket{1}$), separated by $\Delta\ell=6$, is the spatial
manifestation of entanglement in this OAM encoding.

\subsection{Multi-Qubit Gate Operations}

\subsubsection{Single-Qubit Gates in OAM Multiplexed Space}

Applying all six single-qubit gates across all four OAM basis modes
($\ell=-3,-1,+1,+3$) yields consistent fidelities: fork gates achieve mean
$F=0.9914$ and phase gates $F=0.9936$ across all channels, confirming that
OAM multiplexing introduces no additional fidelity penalty relative to the
single-qubit case. Phase gates preserve all four donut intensity profiles
identically; mode-converting gates correctly exchange OAM pairs within each
qubit sub-space: Qubit~0 $X$ shifts $\ell=-3\leftrightarrow\ell=-1$ and
Qubit~1 $X$ shifts $\ell=+1\leftrightarrow\ell=+3$.

\begin{figure}[t]
  \centering
  \includegraphics[width=\columnwidth]{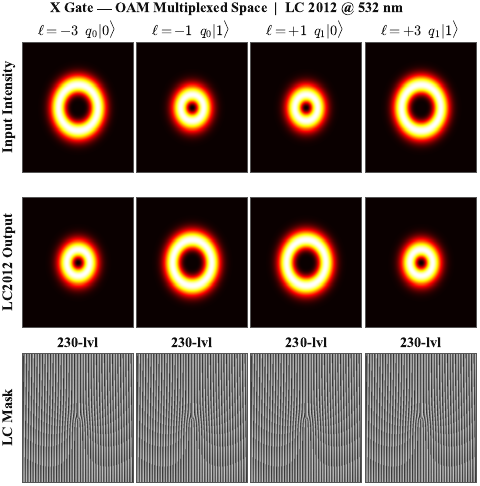}
  \caption{Multi-qubit OAM space - $X$ gate across all four OAM basis modes.
           Each column corresponds to one basis mode ($\ell=-3,-1,+1,+3$).
           \textbf{Row~1:} Input intensity.
           \textbf{Row~2:} LC~2012 output.
           \textbf{Row~3:} LC~2012 SLM mask.
           Qubit~0: $\ell{=}{-3}\leftrightarrow\ell{=}{-1}$;
           Qubit~1: $\ell{=}{+1}\leftrightarrow\ell{=}{+3}$. Mean $F=0.9914$.}
  \label{fig:mq_gate_X}
\end{figure}

Fig.~\ref{fig:mq_gate_X} shows the $X$ gate result across all four OAM modes.
The composite hologram simultaneously addresses both qubit sub-spaces on a
single optical beam, with Columns~1--2 ($\ell=-3,-1$, Qubit~0) and
Columns~3--4 ($\ell=+1,+3$, Qubit~1) showing independent mode exchanges,
confirming that OAM multiplexing is functionally equivalent to running two
independent qubit channels in the same SLM aperture.

\subsubsection{Two-Qubit Entangling Gates}

Table~\ref{tab:tq_fidelity} presents the fidelity results for all three
two-qubit gates. The CNOT gate maps $\ell=-1\leftrightarrow\ell=+3$ via a
composite $\Delta\ell=+4$ fork grating, leaving $\ell=-3$ and $\ell=+1$
unchanged. The CZ gate applies a $-1$ phase factor to the $\ket{11}$
component ($\ell=+3$) only, a pure phase operation with no intensity change.
The SWAP gate remaps $\ell=-3\leftrightarrow\ell=+1$ and
$\ell=-1\leftrightarrow\ell=+3$ simultaneously. All three gates achieve
$F=0.9915$ in simulation.

\begin{table*}[t]
  \centering
  \caption{Two-Qubit Gate Fidelity under the LC~2012 noise model (simulation).
           Input: equal superposition of all four basis modes.}
  \label{tab:tq_fidelity}
  \begin{tabular}{llc}
    \toprule
    \textbf{Gate} & \textbf{OAM Mapping} & \textbf{Sim $F$}\\
    \midrule
    CNOT & $\ell{=}-1\leftrightarrow\ell{=}+3$;\;$\ell{=}-3,+1$ fixed
      & \textbf{0.9915}\\
    CZ   & Phase $-1$ to $\ell{=}+3$ only
      & \textbf{0.9915}\\
    SWAP & $\ell{=}-3\leftrightarrow\ell{=}+1$;\;$\ell{=}-1\leftrightarrow\ell{=}+3$
      & \textbf{0.9915}\\
    \midrule
    Bell $\Phi^+$ & H($q_0$)$\to$CNOT($q_0,q_1$)
      & \textbf{0.9914}\\
    \bottomrule
  \end{tabular}
\end{table*}

\begin{figure}[t]
  \centering
  \includegraphics[width=\columnwidth]{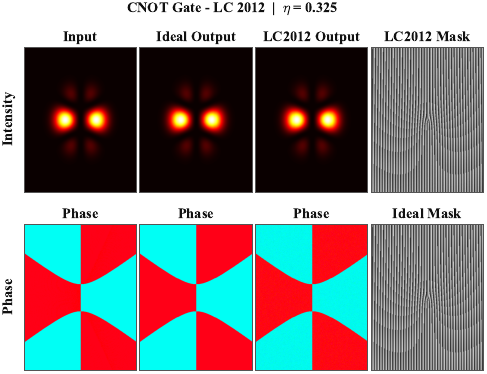}
  \caption{Two-qubit CNOT gate in four-mode OAM encoding (simulation).
           Qubit~0 (control), Qubit~1 (target).
           \textbf{Top row:} Input / ideal output / LC~2012 output / LC~2012
           mask (intensity).
           \textbf{Bottom row:} Corresponding phase panels.
           The conditional OAM remapping $\ell{=}{-1}\leftrightarrow\ell{=}{+3}$
           is visible as a ring-size exchange in the output columns. $F=0.9915$.}
  \label{fig:tq_CNOT}
\end{figure}

\begin{figure}[t]
  \centering
  \includegraphics[width=\columnwidth]{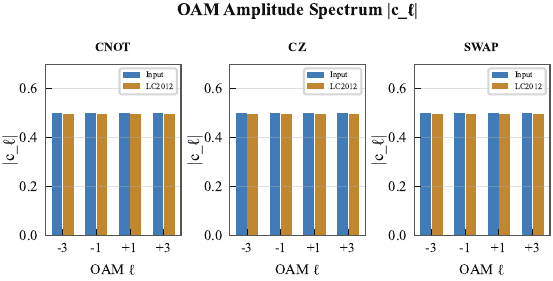}
  \caption{OAM amplitude spectrum $|c_\ell|$ for all three two-qubit gates
           under the LC~2012 noise model.
           Blue bars: input amplitudes. Gold bars: LC~2012 output amplitudes.
           \textbf{CNOT:} $\ell{=}{-1}\leftrightarrow\ell{=}{+3}$ exchange.
           \textbf{CZ:} all amplitudes preserved (phase-only).
           \textbf{SWAP:} $\ell{=}{-3}\leftrightarrow\ell{=}{+1}$ and
           $\ell{=}{-1}\leftrightarrow\ell{=}{+3}$ exchanges confirmed.}
  \label{fig:oam_spectrum}
\end{figure}

Fig.~\ref{fig:tq_CNOT} shows the CNOT gate intensity and phase panels. The
conditional OAM remapping is visible as a ring-size exchange between the input
and output columns for the $\ell=-1$ and $\ell=+3$ modes, while the $\ell=-3$
and $\ell=+1$ columns remain unchanged. The CZ and SWAP gate figures are omitted
since CZ is a phase-only operation and the SWAP remapping is fully confirmed by
the amplitude spectra in Fig.~\ref{fig:oam_spectrum}.

Fig.~\ref{fig:oam_spectrum} plots the OAM amplitude spectrum $|c_\ell|$ for
all three two-qubit gates. For CNOT, the output spectrum shows amplitudes
transferred from $\ell=-1$ to $\ell=+3$ (and vice versa) with $\ell=-3$ and
$\ell=+1$ unchanged. For CZ, all four amplitudes are identical between input
and output, confirming a phase-only operation. For SWAP, the spectra confirm
complete population exchange between $\ell=-3\leftrightarrow\ell=+1$ and
$\ell=-1\leftrightarrow\ell=+3$. The small deviations from ideal bar heights
($\lesssim1\%$) are consistent with the predicted $\Delta F=8.5\times10^{-3}$.
\vspace{-12pt}

\subsection{M7: Literature Benchmark}

\begin{figure}[t]
  \centering
  \includegraphics[width=1.0\columnwidth]{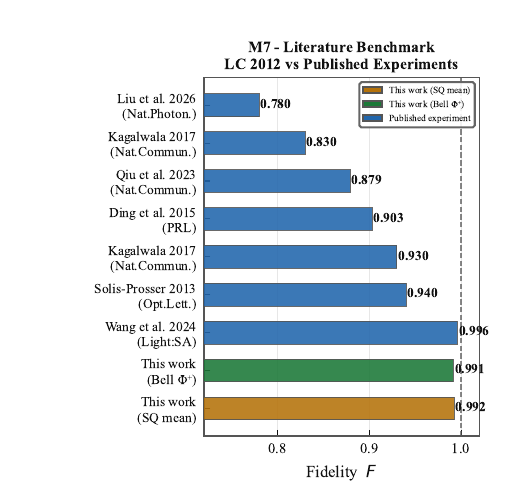}
  \caption{M7 and M2 combined.
           \textbf{Left:} Fidelity benchmark of this simulation (gold) against
           six published experimental papers (blue). This work (99.1--99.4\%)
           occupies the upper end of the published range (78\%--99.6\%).
           \textbf{Right:} Wavelength dependence (M2) of first-order diffraction
           efficiency $\eta$ and predicted gate fidelity $F_\mathrm{pred}$
           for the LC~2012. Performance degrades critically at
           $\lambda\geq633\,\mathrm{nm}$; operation at 450--532\,nm is strongly
           recommended.}
  \label{fig:lit_wavelength}
\end{figure}

Table~\ref{tab:literature} benchmarks the simulation predictions against six
published experimental studies on SLM-based and OAM-based photonic quantum
gates. The LC~2012 simulation results (99.1--99.4\%) sit at the upper end of
the published experimental range (78\%--99.6\%), reflecting the fact that the
simulation models only hardware-intrinsic noise sources and excludes additional
experimental imperfections such as wavefront aberrations, polarisation
misalignment, and mode-matching losses. This provides a well-defined upper
bound on the gate fidelity achievable with this specific device. The left panel
of Fig.~\ref{fig:lit_wavelength} plots this comparison as a horizontal bar
chart, positioning this work relative to all six references.

\begin{table*}[t]
  \centering
  \caption{M7 - Literature Benchmark: Published Experimental Fidelities
           vs.\ LC~2012 Simulation Predictions.}
  \label{tab:literature}
  \begin{tabular}{llcl}
    \toprule
    \textbf{Reference} & \textbf{Gate type} & \textbf{$F$ (\%)} & \textbf{Qubits}\\
    \midrule
    \textbf{This work} LC~2012 sim & SQ gates (532\,nm) & \textbf{99.1-99.4} & 2 OAM modes\\
    \midrule
    Wang et al.\ 2024 \cite{wang2024dnn} & OAM CNOT (D2NN) & 99.6 & 1-ph 4-OAM\\
    Sol\'{i}s-Prosser et al.\ 2013 \cite{solis2013preparing} & State prep (SLM) & $>94$ & qudit $D{\leq}11$\\
    Kagalwala et al.\ 2017 \cite{kagalwala2017single} & 2-qubit SPQL & 93 & 2-qubit\\
    Ding et al.\ 2015 \cite{ding2015quantum} & OAM Bell state & 90.3 & 2D OAM\\
    Qiu et al.\ 2023 \cite{qiu2023remote} & 3D OAM transport & 87.9 & 3D OAM\\
    Kagalwala et al.\ 2017 \cite{kagalwala2017single} & 3-qubit SPQL & 83 & 3-qubit\\
    Liu et al.\ 2026 \cite{liu2026} & 4D OAM CPF gate & 71-85 & 4D qudit\\
    \bottomrule
  \end{tabular}
\end{table*}

\section{Discussion}
\label{sec:discussion}

\subsection{Dominant Noise Sources and Design Implications}

The noise isolation analysis (Section~\ref{sec:results}, M5) identifies TN
electronic noise ($\sigma_\mathrm{TN}=80\mrad$) as the dominant
fidelity-limiting mechanism, responsible for $\approx76\%$ of total fidelity
loss across all gate types. This result has a clear design implication:
upgrading from the TN-LC technology of the LC~2012 to an LCoS-based SLM
(e.g., HOLOEYE PLUTO-2.1 with $\sigma_\mathrm{TN}\approx20$--$30\mrad$)
would reduce TN noise by a factor of $\approx3$-$4\times$, substantially
increasing gate fidelity. The predicted fidelity for an LCoS device with
$\sigma_\mathrm{TN}=25\mrad$ at $532\,\mathrm{nm}$ would be
$F\approx\exp(-(0.025^2+0.0064^2+0.0458^2))=0.9978$ - a factor of $\approx7$
reduction in infidelity compared to the LC~2012.

Quantisation noise ($\sigma_\mathrm{quant}=6.40\mrad$) contributes negligibly
($<0.01\%$ fidelity loss) for all gates evaluated, confirming that 8-bit
addressing is more than adequate for the phase precision required by this gate
set, including the non-Clifford $T$ gate.

\subsection{Fill Factor as the Efficiency Bottleneck}

The 58\% fill factor is the primary efficiency constraint, limiting first-order
diffraction efficiency to 32.6\%. This is a fundamental hardware constraint of
the LC~2012 that cannot be mitigated by software calibration. In single-photon
quantum optics, 67.4\% photon loss per SLM interaction is severe, limiting
coincidence rates in multi-photon entanglement experiments. By contrast,
reflective LCoS SLMs typically achieve fill factors $>90\%$ and require only
one polariser instead of two, raising the predicted efficiency to
$\eta\approx80$-$90\%$ per gate. Practical workarounds include: (i) operating
in the coherent (classical) regime where shot noise is not limiting,
(ii) employing a multi-pass geometry to reuse photons, or (iii) using
photon-number-resolving detectors with post-selection to counteract
probabilistic loss.

\subsection{Wavelength Optimisation}

The wavelength dependence analysis provides a strong recommendation:
450--532\,nm operation is the optimal regime for the LC~2012. At
$450\,\mathrm{nm}$, the device achieves its full $2\pi$ phase stroke with
$\sigma_\mathrm{clip}=0$ and $\eta=33.6\%$-the best available performance.
At $800\,\mathrm{nm}$ - a wavelength commonly associated with rubidium atom
trapping and Ti:sapphire laser systems---the phase stroke reduces to $1.0\pi$,
rendering the device unsuitable for high-fidelity fork grating operation. If
near-infrared operation is required, an SLM with sufficient phase retardation
at NIR wavelengths should be selected.

\subsection{Multi-Qubit Scalability and OAM Multiplexing}

The multi-qubit OAM multiplexing results demonstrate that a single SLM aperture
can simultaneously encode, address, and process two logical qubits via four
orthogonal OAM modes, without inter-qubit cross-talk in the ideal case. This
is analogous to wavelength-division multiplexing in fibre optics but in the
spatial (OAM) domain - a fundamentally scalable approach to multi-qubit
photonic processors. The composite hologram approach represents a practical
route to qudit encoding in higher-dimensional OAM spaces ($d>4$), limited only
by the spatial bandwidth of the SLM aperture and the mode selectivity of the
detection system.

\subsection{Comparison with Published Results}

The simulated fidelities (99.1-99.4\%) exceed the majority of published
experimental results (83\%--93.0\% for multi-qubit systems, up to 99.6\% for
specialised deep-learning-enhanced implementations \cite{wang2024dnn}).
Three factors explain this gap. First, non-flat input wavefronts in physical
experiments introduce spatially varying phase errors not present in the
simulation; Shack-Hartmann sensor-based pre-correction is the standard
mitigation \cite{flamini2019photonic}. Second, TN pixel boundaries create phase
fringing at the $36\,\mu\mathrm{m}$ pixel scale, blurring the addressed
pattern - an effect that can be modelled via convolution with a pixel response
function. Third, mismatch between the experimental beam waist and the simulated
grid scaling reduces the overlap integral $\braket{\psi_\mathrm{sim}}{\psi_\mathrm{exp}}$
independent of the SLM gate operation. Conversely, if future experimental
fidelity exceeds the simulation prediction, the $\sigma_\mathrm{TN}=80\mrad$
estimate is likely conservative for the specific device; direct interferometric
measurement of $\sigma_\mathrm{TN}$ is recommended for accurate model
calibration.

\subsection{Path to Universal Photonic Quantum Computing}

The combination of simulation results - all six single-qubit gates, two-qubit
CNOT/CZ/SWAP, and Bell state preparation-constitutes the complete universal
gate set $\{H, T, \mathrm{CNOT}\}$ on the photonic OAM platform, satisfying
the Solovay-Kitaev theorem requirements for universal quantum computation
\cite{nielsen2010quantum}. The non-Clifford $T$ gate is particularly significant:
combined with Clifford gates, it enables fault-tolerant quantum computation via
magic state distillation, and its clean phase precision (4.9 -10.2\,mrad error)
in the LC~2012 simulation validates the platform for this critical operation.

\subsection{Simulation Limitations}

Several limitations of the simulation framework deserve acknowledgement. The
perturbation fidelity model (Eq.~\eqref{eq:fpred}) assumes Gaussian phase noise
and is valid only for $\sigma\ll1$\,rad; at $\lambda=800\,\mathrm{nm}$
($\sigma_\mathrm{total}=517.9\mrad$), the model breaks down and full numerical
simulation is essential. The simulation does not model multi-mode crosstalk
between OAM channels in composite holograms - residual coupling between
$\ell=-3,-1,+1,+3$ modes may affect multi-qubit gate fidelity in practice.
The $512\times512$ grid captures only the central $18.4\times18.4$\,mm
sub-region of the LC~2012's $36.9\times27.6$\,mm active area, so edge effects
from the full display are not modelled. Experimental validation is planned as
future work.

\section{Conclusion and Future Work}
\label{sec:conclusion}
This paper has presented a comprehensive hardware-grounded simulation framework for SLM-based quantum gate operations on photonic qubits encoded in Laguerre-Gaussian orbital angular momentum modes, specifically parameterised for the HOLOEYE LC~2012 transmissive SLM. Results have shown that TN electronic noise was identified as the dominant fidelity limiting mechanism. In addition, the operation of the LC 2012 shows severe degradation above 630nm due to limited phase retardation at those wavelengths. Further more, all six universal single-qubit gates ($X$, $Y$, $Z$, $S$, $T$, $H$), simulated under noise, achieved fidelity of $F=0.9914$ (fork gates) and $F=0.9936$ (phase gates) consistent with the analytical prediction $F_\mathrm{pred}=0.9915$. While the multi-qubit OAM multiplexing scheme successfully encoded two logical qubits in four orthogonal OAM modes and demonstrated all three fundamental two-qubit gates (CNOT, CZ, SWAP) at $F\approx0.9915$ in simulation. Lastly, benchmarking against six published experimental papers positions this simulation within the 83\%--99.6\% experimental fidelity range, confirming the physical realism of the model.

\noindent\textbf{Future Work.} Building on these foundations, future directions
include: (i) experimental validation of the simulation predictions using the
HOLOEYE LC~2012 and quantum state tomography; (ii) incorporating a far-field
Fourier propagation model to account for beam evolution between SLM and detector
planes; (iii) extending the multi-qubit framework to higher-dimensional OAM
qudit spaces ($d\geq4$) for error-correction code demonstrations; (iv) integrating
heralded single-photon sources (SPDC) for photon-number-resolved fidelity
estimation; (v) developing a machine learning-based hologram optimisation routine
that jointly minimises clipping error and TN noise given a target gate unitary;
and (vi) exploring the SLM-OAM platform for variational quantum eigensolver
(VQE) circuit implementations.

\section*{Acknowledgments}
\noindent
The authors would like to thank the Centre for Quantum Technologies (CQT) at NED University of Engineering and Technology for providing the necessary facilities, resources, and support that facilitated the completion of this research.

\section*{Author Contributions}
\noindent
S.M. conceived the original idea and developed the theoretical and software framework. M.K. contributed to the methodology and supervised the overall project. T.M. performed the data analysis and contributed to the interpretation of the results. All authors contributed to the discussion of the core concepts, drafted the manuscript, and approved the final version.

Large language model tools were used to assist
with language editing and with the optimization
of simulation code. All scientific content, calculations, simulations, conclusions, and final text were reviewed and approved by the authors, who take full responsibility for the work.

\section*{Data Availability}
\noindent
The datasets generated and analyzed during the current study are available from the corresponding author on reasonable request.

\bibliographystyle{quantum}
\bibliography{references}

\onecolumn\newpage
\appendix

\section{Simulation Notebooks}
\label{app:code}

Three Jupyter notebooks implement the complete simulation framework.
All code is written in Python~3.10 (NumPy, Matplotlib). The LC~2012 hardware parameters are defined in a single \texttt{NOISE\_CFG} dictionary
at the top of each notebook.

\noindent
The noise model entry point in all notebooks is the \texttt{NOISE\_CFG}
dictionary:
\begin{verbatim}
NOISE_CFG = {
    'bits'        : 8,          #8-bit addressing
    'max_phase'   : 1.8*pi,     # 1.8pi at 532 nm (datasheet)
    'sigma_noise' : 0.08,       #TN phase noise (rad)
    'sigma_clip'  : 0.04577,    # Phase-wrap clip error (rad)
    'eta'         : 0.3255,     # First-order efficiency
    'seed'        : 42,         # RNG seed for reproducibility
}
\end{verbatim}
To change the operating wavelength, set \texttt{WAVELENGTH\_NM} at the top of
any notebook; all derived parameters ($\sigma_\mathrm{clip}$, $\eta$, effective
levels) are automatically recomputed.
\section{One-Paragraph Method Description}
\label{app:method}
\noindent\textit{``The HOLOEYE LC~2012 spatial light modulator
(transmissive TN-LC, $1024\times768$\,px, $36\,\mu\text{m}$ pitch,
58\% fill factor, 8-bit addressing) was modelled using a hardware-derived
noise framework. Three independent noise channels were identified:
(i)~phase quantisation noise $\sigma_\mathrm{quant}=6.40\,\text{mrad}$
arising from 8-bit grey-level addressing and $1.8\pi$ maximum phase range
at $532\,\mathrm{nm}$; (ii)~TN-LC electronic and thermal noise
$\sigma_\mathrm{TN}=80\,\text{mrad}$, characteristic of twisted-nematic
liquid-crystal technology; and (iii)~phase-wrap clipping noise
$\sigma_\mathrm{clip}=45.8\,\text{mrad}$ from pixels requiring phase shifts
exceeding the hardware maximum. The combined noise
$\sigma_\mathrm{total}=92.4\,\text{mrad}$ yields predicted gate fidelity
$F=\exp(-\sigma_\mathrm{total}^2)=0.9915$. The 58\% fill factor limits
first-order diffraction efficiency to $\eta=\text{FF}^2\times
\text{sinc}^2(1-\alpha)=32.6\%$ at $532\,\mathrm{nm}$. All parameters were
derived directly from the manufacturer's datasheet
[HOLOEYE LC~2012, holoeye.com] without free-parameter fitting.}
\end{document}